\newif\ifmnras
\mnrastrue
\ifmnras
\documentclass[useAMS,usenatbib,usedcolumn]{mnras}
\else
\documentclass[apj,numberedappendix]{emulateapj}
\fi
\usepackage{graphics,epsf}
\usepackage{amsmath}                
\usepackage{amsfonts}               
\usepackage{amssymb}                
\usepackage{epsfig}                 
\usepackage{graphicx}
\usepackage{float}

\newcommand{\km}{{~\rm km}}
\newcommand{\s}{{~\rm s}}

\newcommand{\yr}{{~\rm yr}}

\newcommand{\ourlongtitle}{{Type~IIb supernovae by the grazing envelope evolution}}
\newcommand{\ourshorttitle}{{SNe~IIb by the GEE}}
    
\ifmnras
\else
\newcommand{\nar}{{~\rm New Astronomy Reviews}}
\newcommand{\na}{{~\rm New Astronomy}}
\newcommand{\pasa}{{~\rm Publications of the Astronomical Society of Australia}}
\fi
   
\ifmnras
\title[\ourshorttitle]{\ourlongtitle}
\author[B. V. Naiman, E. Sabach, A. Gilkis, N. Soker]{Binyamin V. Naiman$^{1}$
Efrat Sabach$^{1}$\thanks{Contact e-mail: \href{efrats@physics.technion.ac.il}{efrats@physics.technion.ac.il}}, Avishai Gilkis$^{2}$\thanks{Contact e-mail: \href{agilkis@ast.cam.ac.uk}{agilkis@ast.cam.ac.uk}}, Noam Soker$^{1,3}$\thanks{Contact e-mail: \href{soker@physics.technion.ac.il}{soker@physics.technion.ac.il}}
\\
$^{1}$ Department of Physics, Technion -- Israel Institute of Technology, Haifa 3200003, Israel \\
$^{2}$ Institute of Astronomy, University of Cambridge, Madingley Road, Cambridge, CB3 0HA, UK  \\
$^{3}$ Guangdong Technion Israel Institute of Technology, Shantou, Guangdong Province 515069, China
}
\fi

\begin{document}

\ifmnras
\label{firstpage}
\pagerange{\pageref{firstpage}--\pageref{lastpage}} \pubyear{2018}
\maketitle
\else
\title{\ourlongtitle}
\author{Binyamin V. Naiman\altaffilmark{1}, Efrat Sabach\altaffilmark{1}, Avishai Gilkis\altaffilmark{2} \& Noam Soker\altaffilmark{1,3}}
\altaffiltext{1}{Department of Physics, Technion -- Israel Institute of Technology, Haifa
3200003, Israel; bnaiman@campus.technion.ac.il; XXXXXX@XXXXXX.technion.ac.il; soker@physics.technion.ac.il}
\altaffiltext{2}{Institute of Astronomy, University of Cambridge, Madingley Road, Cambridge, CB3 0HA, UK; agilkis@ast.cam.ac.uk}
\altaffiltext{3}{Guangdong Technion Israel Institute of Technology, Shantou, Guangdong Province 515069, China}
\fi

\begin{abstract}
We simulate the evolution of binary systems with a massive primary star of $15 M_\odot$ where we introduce an enhanced mass loss due to jets that the secondary star might launch, and find that in many cases the enhanced mass loss brings the binary system to experience the grazing envelope evolution (GEE) and form a progenitor of Type~IIb supernova (SN~IIb). The jets, the Roche lobe overflow (RLOF), and a final stellar wind remove most of the hydrogen-rich envelope, leaving a blue-compact SN~IIb progenitor. In many cases without this jet-driven mass loss the system enters a common envelope evolution (CEE) and does not form a SN~IIb progenitor. We use the stellar evolutionary code \textsc{MESA~binary} and mimic the jet-driven mass loss with a simple prescription and some free parameters. Our results show that the jet-driven mass loss, that some systems have during the GEE, increases the parameter space for stellar binary systems to form SN~IIb progenitors. We estimate that the binary evolution channel with GEE contributes about a quarter of all SNe~IIb, about equal to the contribution of each of the  other three channels, binary evolution without a GEE, fatal CEE (where the secondary star merges with the core of the giant primary star), and the single star channel.
\ifmnras
\else
\fi
\end{abstract}

\ifmnras
\fi

\section{INTRODUCTION}
\label{sec:intro}

\subsection{Type~IIb supernovae (SNe~IIb)}
\label{subsec:SNIIb}

Supernovae IIb (SNe~IIb) are classified as core collapse supernovae (CCSNe) that have strong hydrogen lines at early times, days after explosion, which later substantially weaken and even disappear. The weakening of the hydrogen lines results from a low mass hydrogen-rich envelope of the SN~IIb progenitor. 
This behaviour implies that the progenitor of the CCSN has a very small hydrogen mass at the time of explosion, $M_{\rm H} \simeq 0.03-0.5 M_\odot$ (e.g., \citealt{Woosleyetal1994, Meynetetal2015, Yoonetal2017}), {{{{ with a possible lower mass of even down to $ M_{\rm H} \simeq 0.001 M_\odot$ \citep{Dessartetal2011, Eldridgeetal2018}. }}}} In their population synthesis study \cite{Sravanetal2018} take the hydrogen-rich envelope of the progenitor at the onset of explosion to have a mass of $0.01 M_\odot \le M_{\rm H,env} \le 1 M_\odot$. 

SNe~IIb amount to about $f_{\rm IIb} \simeq 11 \%$ of all CCSNe \citep{Smithetal2011, Shivversetal2017, Grauretal2017b}. \cite{Grauretal2017b} find that the relative rates of SNe~IIb do not depend much on the mass of their host galaxies. \cite{Sravanetal2018} take $f_{\rm IIb,H} \simeq 10-12 \%$ in high metallicity stellar populations and $f_{\rm IIb, L} \simeq 20 \%$ in low metallicity populations.   

There is observational support for the binary scenario for the formation of SNe~IIb. 
\cite{Kilpatricketal2017} fit a binary model for the progenitor of SN 2016gkg {{{{ (reported by \citealt{Berstenetal2018}) }}}} with an initial period of $P_i= 1000~$days, and initial stellar masses of $M_{\rm 1,i}=15M_\odot$ and $M_{\rm 2,i}=1.5 M_\odot$. The pre-explosion primary mass in their fitting is $M_{1,f}=5.2 M_\odot$. 
\cite{Alderingetal1994} deduce from the photometry of the SN~IIb 1993J that it better fits a binary progenitor, as suggested by \cite{Podsiadlowskietal1993}.
{{{{ Later, \cite{Maundetal2004} observed the companion. }}}}
\cite{Foxetal2014} argue that the flattened circumstellar matter around SN 1993J \citep{Mathesonetal2000} supports a binary progenitor. 
\cite{Fremlingetal2019} study the SN~IIb ZTF18aalrxas, and find its hydrogen mass to be $\approx 0.15 M_\odot$, and argue that the zero age main sequence (ZAMS) mass of its progenitor was about $12 M_\odot$. They further find massive CSM, and argue that only a binary interaction can explain all these properties. 
\cite{Soker2017} takes a mass loss in a flat disk or ring to support the grazing envelope evolution (GEE) route (section \ref{subsec:GEE} below), as such a structure is found in post-asymptotic giant branch intermediate binaries (post-AGBIBs; e.g., \citealt{Kastneretal2010, VanWinckel2017}), and the main sequence companion in some post-AGBIBs are observed to launch jets (e.g., \citealt{Wittetal2009, Gorlovaetal2012, Thomasetal2013, Gorlovaetal2015, VanWinckel2017b}). The companion orbits outside but close to the post AGB star. 

\cite{Claeys2011} who expand the work of \cite{StancliffeEldridge2009} find that  binary evolution predicts only $\approx 0.6 \%$ of all CCSNe to be SN~IIb, much lower than the observed fraction. They could increase this fraction if 
they consider low angular momentum loss from the binary system and low accretion efficiencies by the companion, such that the specific angular momentum lost in the outflow is smaller than that of the binary system. \cite{OuchiMaeda2017} also share the conclusion of a large mass loss fraction.
\cite{Soker2017} attributes the efficient mass removal from the binary system to jets that the companion star launches as it accretes mass from the SN progenitor. \cite{Soker2017} argues that the jets in the GEE scenario both remove mass from the primary stellar progenitor with relatively low specific angular momentum, and limit mass accretion onto the companion itself.

In their very recent population synthesis study \cite{Sravanetal2018} find that single and binary progenitors contribute about equally to the population of SNe~IIb. However, they fall short of explaining the rate of SNe~IIb by a factor of more than 
3 (also \citealt{Sravan2016}). 
Winds that are weaker than usually assumed remove less hydrogen-rich envelope gas after the end of the mass transfer process, and by that can reduce the discrepancy with observations \citep{Gilkisetal2019}. 

The main evolutionary channel that \cite{Sravanetal2018} consider is Roche lobe overflow (RLOF) mass transfer.  
We assume that the vast majority, and possibly all, of SNe~IIb are a result of binary interaction. To boost the number of binary progenitors of SNe~IIb we include two more evolutionary channels. The first one is the GEE route, as proposed by \cite{Soker2017} and which is the subject of the present study. The second evolutionary route is that where a main sequence companion ejects all the original hydrogen-rich envelope, and then is destroyed on to the core of the massive star. The secondary star becomes the new low-mass hydrogen-rich envelope of the massive star \citep{Lohevetal2019}. \cite{Lohevetal2019} suggest this fatal common envelope evolution (FCEE) scenario to explain the SN~IIb Cassiopeia~A.  

A useful classification of SNe~IIb progenitors is to extended progenitors (i.e., red supergiants \citealt{ChevalierSoderberg2010}) and compact progenitors.  \cite{Yoonetal2017} discuss blue progenitors, yellow supergiant progenitors, and red supergiant progenitors, and their formation via RLOF. The first two groups are compact and have little hydrogen mass at explosion, $M_{\rm H} \la 0.15 M_\odot$. 
The hydrogen mass at explosion of the red supergiant progenitors is $M_{\rm H} \ga 0.15M_\odot$. Both stable and unstable mass transfer can form compact progenitors of SNe~IIb, that make most of the SNe~IIb. The GEE cases that we simulate in the present study lead to the formation of blue-compact SN~IIb progenitors, as post-GEE winds remove most of the hydrogen that is left after the GEE. \cite{Yoonetal2017} already noted that post-RLOF winds are efficient in removing most of the left-over hydrogen (see also \citealt{Gilkisetal2019}). 
Winds in higher metallicity populations are more efficient in removing mass, therefore leading to a higher ratio of SNe~Ib to SNe~IIb. 
   
Before we turn to mimic the GEE in simulations, we briefly describe the basic properties of the GEE and the general motivation to introduce the GEE into binary stellar evolution. \cite{Soker2017} presents in more details some of the properties of the GEE that are relevant to the formation of SNe~IIb. 

\subsection{The grazing envelope evolution (GEE)}
\label{subsec:GEE}
There are several results that motivate the introduction of the GEE. (1) The observations of post-AGBIBs, where a secondary star is close but outside the envelope of a post-AGB star (e.g., \citealt{Manicketal2017, Oomenetal2018}) where traditional evolutionary calculations predict no binary systems (e.g., \citealt{Nieetal2012}). 
(2) The observations that the companion in many post-AGBIBs launches jets, even wide jets 
(e.g., \citealt{Thomasetal2013}). 
(3) The failure of most hydrodynamical simulations of the common envelope evolution (CEE) to eject the entire envelope in a consistent and persistent manner (e.g., \citealt{TaamRicker2010, DeMarcoetal2011, Passyetal2012, RickerTaam2012, Nandezetal2014, Ohlmannetal2016, Staffetal2016MN8, NandezIvanova2016, Kuruwitaetal2016, IvanovaNandez2016, Iaconietal2017, DeMarcoIzzard2017, Galavizetal2017, Chamandyetal2019, Reichardtetal2019}). These simulations might hint on the need for an extra energy source to eject the envelope. {{{{ The `extra' refers to an energy in addition to the gravitational energy that the inspiral binary system releases. Over a very long evolutionary time, such an energy source might be the nuclear burning in the giant core. This causes a mass loss even in single stars, but in the case of a CEE the perturbed giant envelope facilitates dust formation which in turn leads to an enhanced mass loss rate by radiation pressure on dust (\citealt{Soker2004, GlanzPerets2018}). However, here we seek a more rapid mass loss rate.  }}}}

The GEE posits that this extra energy source is the gravitational energy that is released by mass that the more compact companion accretes, and that jets carry this energy to the ambient gas (\citealt{Soker2016Rev} for a review).
{{{{ We note that in that case there can be enhanced mass loss from the giant envelope even if the companion does not spiral-in into the giant envelope, unlike in the case of the CEE. }}}}
\cite{BlackmanLucchini2014} suggest that the high momenta in bipolar planetary nebulae indicate that the companion can launch jets in a CEE. 
From the theoretical side, the energy and the high entropy gas that the jets themselves can carry away from the accretion flow allows a high accretion rate (e.g., \citealt{Shiberetal2016, Staffetal2016MN, Chamandyetal2018a}). Without this energy removal the gas would build a high pressure zone near the accreting object. Such a high pressure zone reduces the accretion rate (e.g. \citealt{RickerTaam2012, MacLeodRamirezRuiz2015}). 

In the GEE jets that the more compact secondary star launches as it grazes the envelope of a giant star remove mass from the envelope  \citep{SabachSoker2015, Soker2015, Shiberetal2017, ShiberSoker2018, LopezCamaraetal2019, Shiberetal2019}. The GEE occurs when the jets efficiently remove mass from the giant envelope near the orbit of the companion. Such a mass removal can delay, and even prevent, the full CEE. The interplay between mass loss, mass accretion, and mass removal by jets determines which one of these outcomes takes place.
(1) The system enters a CEE. (2) The orbital separation substantially decreases as the binary system experiences the GEE. (3) The orbital separation does not change by much. (4) The orbital separation somewhat increases. 

We would like to emphasise the differences between the GEE, that we propose as one of the main channels to form SN~IIb progenitors, and the case of a RLOF that is usually discussed in the literature (section \ref{subsec:SNIIb}). 
(1) In the RLOF process the gravity of the companion and the winds remove mass from the giant envelope. Therefore, if the system loses synchronisation or if the giant expands further, the RLOF process by itself would not be able to prevent the system from entering a CEE. In the GEE the extra energy source that the jets supply can remove more envelope mass and in some cases prevent the CEE. The GEE, hence, substantially increases the parameter space for the formation of SNe~IIb. (2) In the RLOF process most of the mass flows through the first Lagrangian point. In the GEE the accretion process on to the companion is a combination of a RLOF and a Bondi-Hoyle-Lyttleton type accretion. (3) In the RLOF process the companion orbits well outside the giant envelope, while in the GEE the companion grazes the giant envelope. We note that on average the orbital separation during the GEE can be smaller than the radius of the giant star since in the vicinity of the secondary star the jets remove envelope mass and the edge of the envelope at the secondary location is smaller.
 
Our goal is to show that the GEE can increase the parameter space for the formation of SNe~IIb. Namely, to show that the jets of the GEE can prevent systems that otherwise would have entered a CEE from entering a CEE, and that the hydrogen mass at core collapse in some of these systems is that expected for SNe~IIb. 
In this, still preliminary, study we mimic the GEE by changing the parameters of mass transfer and mass loss with the \textsc{mesa~binary} code.   
In section \ref{sec:Mimicking} we describe our numerical scheme to mimic the GEE, and in section \ref{sec:results} we present our results. 
We summarise in section \ref{sec:summary}.
 
\section{MIMICKING THE GRAZING ENVELOPE EVOLUTION}
\label{sec:Mimicking}

We use the \textsc{binary} module of the \textsc{mesa} code (Modules for Experiments in Stellar Astrophysics, version 10398; \citealt{Paxtonetal2011, Paxtonetal2013, Paxtonetal2015, Paxtonetal2018}) to follow the evolution of binary systems. Since our goal is to demonstrate that the GEE can extend the binary parameter space for the formation of SNe~IIb, we limit the study to a small number of cases and to circular orbits. We are not yet in a stage that allows us to explore the absolute number of SNe~IIb that result from the GEE channel, because we did not converge yet on the exact scheme to use for jet-driven mass loss. This is the second study of the GEE with \textsc{mesa~binary}, and we differ quite a lot from the scheme used in the first study with \textsc{mesa~binary} \citep{AbuBackeretal2018}. 

{{{{ As in that earlier study, we assume that the mass that the secondary star accretes from the giant envelope has enough specific angular momentum to form an accretion disk around the secondary star. We further assume that the accretion disk launches jets in a similar manner to the way accretion disks launch jets around young stellar objects. }}}}

We evolve a binary system starting with two main sequence stars of ZAMS masses of $M_{\rm 1,i}=15 M_\odot$ (the primary mass donor star) and a secondary star with an initial mass of $M_{\rm 2,i}=2.5 M_\odot$ in most cases, while in some cases $M_{\rm 2,i}=2.0 M_\odot$ and $M_{\rm 2,i}=3.0 M_\odot$. The initial metallicity of the primary star is $Z=0.019$ and its initial rotation velocity is zero. We treat the secondary star as a point mass and do not follow its evolution or change of structure as a result of mass accretion (see \citealt{AbuBackeretal2018}). We set the initial orbital separation to be in the range of $a=800-1200 R_\odot$.

The system evolves according to mass loss, mass transfer, and tidal interaction (from \citealt{Hut1981}, with the timescales of \citealt{Hurley2002} for convective envelopes). These interactions can spin-up the primary star. In that case the numerical code treats rotation according to the `shellular approximation', where the angular velocity $\omega$ is assumed to be constant for isobars (e.g., \citealt{Meynet1997}). 

If the stars achieve contact, i.e., the separation equals the sum of their radii,
\begin{equation}
a=R_1+R_2, 
\label{eq:aR1R2}
\end{equation}
where in our simulations here $R_2=0$, in most cases we terminate evolution. In some runs this condition is never met, and the evolution is terminated when the primary star almost reaches core collapse. In some cases we do follow the system after the companion enters the envelope of the giant star, i.e., the system enters a CEE, although the calculation is much less accurate in that case. 
When the companion enters the envelope the simple tidal formulae do not hold any more as the envelope is highly distorted (see simulations cited in section \ref{subsec:GEE}). As well, the accretion processes involves now the Bondi-Hoyle-Lyttleton type flow, and the formulae of the RLOF are not accurate.  

In some simulations we do not introduce jets even when the two stars enter a CEE. These serve for comparison. In other cases we do introduce jets according to the assumption of the GEE (section \ref{subsec:GEE}). 
In both classes of simulations the mass transfer rate due to RLOF, $\dot M_{\rm KR}$, is according to \cite{Kolb1990}. In most cases we assume that only a fraction $f_{\rm acc,RL}=0.3$ of this mass is accreted, and the rest is lost by the system, a fraction of $f_{\rm L,RL,1}$ is lost from the primary giant star, and a fraction of $f_{\rm L,RL,2}$ is lost by the secondary (the figures below present mainly cases with $f_{\rm L,RL,1}=0$ and $f_{\rm L,RL,2}=0.7$). The relation $f_{\rm acc,RL} + f_{\rm L,RL,1}+f_{\rm L,RL,2}=1$ holds. In two runs that we present in section \ref{subsec:Cases} we take $f_{\rm acc,RL}=1$. 
  
Jets that the secondary star launches remove mass, according to our assumption, from the primary envelope and from the acceleration zone of its wind \citep{Hilleletal2020}. We assume that jets remove mass when the orbital separation is   
\begin{equation}
a<f_\mathrm{GEE} \left(R_1+R_2\right),
\label{eq:fGEE}
\end{equation}
where here $R_2=0$, and $f_\mathrm{GEE}$ is the jet-activity separation factor for which we take values of $f_\mathrm{GEE}=1.1- 1.5$ in the different runs. When the condition of equation (\ref{eq:fGEE}) is met, then in addition to the mass transfer $\dot{M}_{\rm KR}$ and mass loss rates of $f_{\rm L,RL,1}\dot M_{\rm KR}$ and $f_{\rm L,RL,2}\dot M_{\rm KR}$ from the primary and secondary stars, respectively, we include extra mass loss resulting from the effect of the assumed jets. The expression for the total jet-driven mass loss rate is
\begin{equation}
\dot M_{\rm L,jet} = f_{\rm jet} \dot M_{\rm KR}  
\frac{f_\mathrm{GEE} - a/R_1} {f_\mathrm{GEE} - 1} ; 
\quad \frac {a}{R_1}<f_{\rm GEE},  
\label{eq:MLJ1}
\end{equation}
where $f_{\rm jet}$ is the jet-driven mass loss factor, and we calculate cases with $f_{\rm jet}=2$ or $f_{\rm jet}=4$.
Half of the mass loss due to jets is from the giant primary star and half from the secondary star.  

We take these values for $f_{\rm jet}$ from the following consideration. The secondary star accretes at a rate of $\dot M_{2\rm , acc}=f_{\rm acc,RL} \dot M_{\rm KR}$, in our simulations. We assume that the secondary launches a fraction of $\eta_j \approx 0.2$ of the accreted mass in jets, and that the jets have a velocity of about the escape velocity from the secondary star, $v_j \simeq 700 \km \s^{-1}$. The escape velocity from the surface of the giant star in our simulations at the relevant time is $v_1 \simeq 100 \km \s^{-1}$. For a maximum efficiency of energy conversion from jets to envelope removal, the jets can remove a mass at a rate of $\dot M_{\rm rem} \approx \eta_j (v_j/v_1)^2 f_{\rm acc,RL} \dot M_{\rm KR}$, which we can scale to read 
\begin{equation}
\frac{ \dot M_{\rm rem}}{\dot M_{\rm KR}} \approx 3
\left( \frac{\eta_j}{0.2} \right) 
\left( \frac{f_{\rm acc,RL}}{0.3} \right)
\left( \frac{v_j}{7v_1} \right)^2 .
\label{eq:fjet}
\end{equation}
This corresponds to $f_{\rm jet} \approx 3$ when the mass removal occurs for $a=R_1$. The highest mass removal by jets that equation (\ref{eq:MLJ1}) gives is when the secondary enters the giant envelope. At that stage the interaction becomes more complicated and the efficiency might increase even more (section \ref{subsec:Cases}).     
  
We found that we need to reduce the time step when the jet activity begins. 
We did so by setting the \textsc{mesa} variable \texttt{varcontrol\char`_target} to $10^{-5}$ (instead of the default value, $10^{-4}$). 

The scheme for mass loss by winds follows \cite{deJager1988} when the effective surface temperature $T_\mathrm{eff}$ is below $10^4\,\mathrm{K}$. For hot phases ($T_\mathrm{eff} \ge 1.1 \times 10^4 \,\mathrm{K}$) we follow \cite{Vink2001} if the surface hydrogen mass fraction $X_\mathrm{s}$ is above $0.4$ or \cite{Vink2017} when $X_\mathrm{s} \le 0.4$. For $1.1\times 10^4\,\mathrm{K}>T_\mathrm{eff} > 10^4\,\mathrm{K}$ we interpolate.

\section{RESULTS}
\label{sec:results}

\subsection{Preventing the CEE and forming SNe~IIb}
\label{subsec:prevent}

Our aim in this subsection is to show that under the assumptions of the GEE, in some cases jets might prevent the system from entering a CEE and by that form a progenitor of a SN~IIb. For that we present the evolution of the orbital separation and masses of a binary system in two cases. In one case we ignore any effects of jets and find that the system enters a CEE. In the other case we introduce an enhanced mass loss rate by jets according to the assumptions of the GEE, and find that the system does not enter a CEE, and that the primary star reaches core collapse when its hydrogen content is that of a SN~IIb. {{{{ We will present and analyse only a number of cases out of the 104 runs that we performed and that we list in Appendix \ref{AppendixA}. }}}}

We first present the evolution of a binary system with an initial circular orbit with a radius of $a_0=1000 R_\odot$, and initial masses of $M_{\rm 1,i}=15 M_\odot$ and $M_{\rm 2,i}=2.5 M_\odot$. The primary starts with no rotation. In run Jet($a_0,f_{\rm GEE},f_{\rm jet}$)=Jet(1000,1.2,4) we turn on the jets according to equation (\ref{eq:MLJ1}) with $f_{\rm GEE}=1.2$ and $f_{\rm jet}=4$. In run NoJet(1000) we do not consider jets. In section \ref{sec:Mimicking} we list the other parameters of the simulations. In all runs presented in sections \ref{subsec:prevent} and \ref{subsec:Numerical} $f_{\rm acc,RL}=0.3$, i.e., the secondary star accretes $30\%$ of the mass that the primary star transfers to it in the RLOF process. 

\begin{figure*}
\centering
\includegraphics[trim=0.0cm 6.cm 0.0cm 9.0cm,clip=true,width=1.0\textwidth]{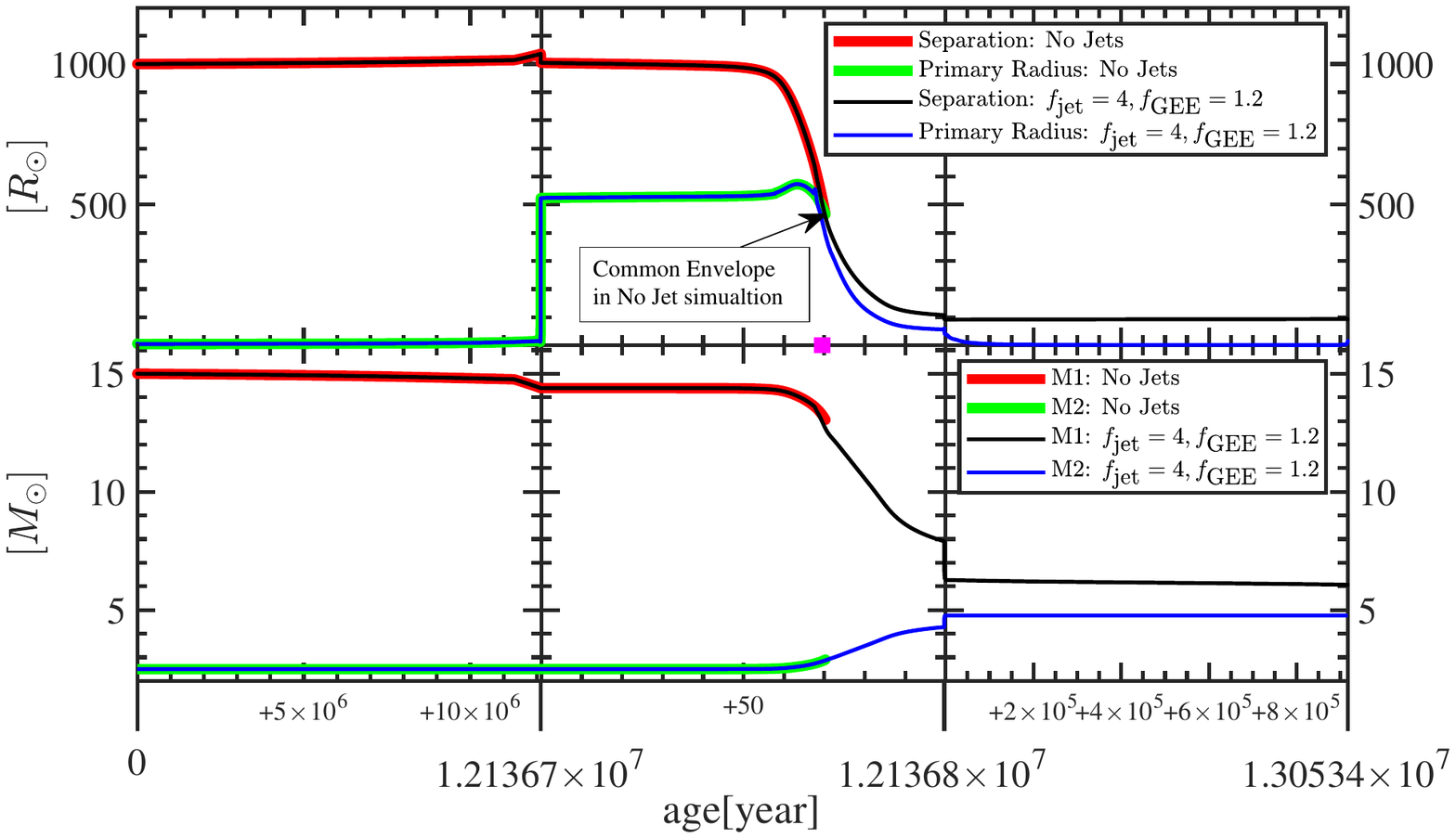}
\vskip -2.0 cm
\caption{Evolution of orbital separation and primary radius (upper panel) and masses (lower panel) from the zero age main sequence to core collapse of the primary star for the Jet($a_0,f_{\rm GEE},f_{\rm jet}$)=Jet(1000,1.2,4) and NoJet(1000) cases. The horizontal time axis has three segments with different timescales. Each lower large number gives the time in years from the zero age main sequence to the respective large tick that marks the beginning of the time segment, while the smaller numbers closer to the axis give the extra time in years from the beginning of the time segment. The thick red line in the upper panel shows the orbital separation for the NoJet(1000) case where jet activity does not take place. The system enters a common envelope phase and the calculation is terminated. The thin black line presents the orbital separation for the Jet(1000,1.2,4) case, where jet activity takes place when $a<f_{\rm GEE} R_1$, with $f_{GEE}=1.2$, according to equation (\ref{eq:MLJ1}). The short thick magenta horizontal line on the boundary between the two panels marks the jets' activity period, about four years. }
\label{fig:Jet1000p1n2p4Ev}
\end{figure*}

We present the evolution of the masses and of the orbital separation with time for the two cases in Fig. \ref{fig:Jet1000p1n2p4Ev}. To present the entire evolution on one graph, we split the horizontal time axis to three segments, each of a different timescale. Initially the stars have a weak interaction between them and the orbital separation does not change much. At about $t=1.213675 \times 10^7 \yr$, the ratio $R_1/a$ becomes large enough for tidal interaction to act fast. Tidal forces transfer orbital angular momentum to the spin of the giant star, and the orbital separation rapidly decreases. Without jets the system enters a CEE as we mark on the figure, where we terminate the evolution (end of thick-red and thick-green lines). 
In the case with jets, as depicted by the thin-black lines, the system does not enter a CEE, and the two stars are detached at explosion (end of graph). Although the jets are active for only four years (thick magenta line on the middle horizontal axis) they manage to prevent the CEE (more in section \ref{subsec:Numerical}).

The jets prevent the CEE by removing mass from the system (section \ref{sec:Mimicking}), a process that acts to increase the orbital separation. If the ratio of orbital separation to primary (giant) radius, $a/R_1$, increases, the effect of the jets decreases (equation \ref{eq:MLJ1}) and tidal forces decrease back this ratio. If the ratio $a/R_1$ decreases then the mass removal rate by jets increases, a process that acts to increase orbital separation. The outcome, as the GEE requires \citep{Soker2015}, is that the orbital separation and primary radius are very close to each other during the GEE, as we see in the upper panel of Fig. \ref{fig:Jet1000p1n2p4Ev} in the last part of the middle time segment. During part of this time jets are active, and in the rest the RLOF keeps the orbital separation and primary radius such that they evolve close to each other. 
The final orbital separation is much smaller than the maximum radius the primary star has achieved. 

During the GEE the secondary star increases its mass, almost doubling it (thin blue line in the lower panel). To prevent the expansion of the secondary star as a result of the high mass accretion, the jets must carry out large amounts of energy and to remove high entropy gas from the vicinity of the secondary star \citep{Shiberetal2016, Chamandyetal2018a}. As \cite{AbuBackeretal2018} already noted, the rotation of the accreting star (which we do not treat here) complicates the accretion process (e.g., \citealt{Kunitomoetal2017}). This process deserves its own study, but at present we note the following (see also also \citealt{AbuBackeretal2018}). 
Any large envelope that the secondary star might inflate will have high entropy. The jets will remove a large fraction of this envelope, i.e., the jets will remove high entropy gas that will limit the expansion of the envelope of the secondary star. Also, it is possible that the dense accretion disk allows the envelope to inflate along the polar directions, but the accretion process through the dense thin accretion disk continues, and therefore the jets are still active. In any case, for our results to hold, it will be necessary to show by simulating the accretion process, that main sequence stars in the mass range $\approx 1.5-3 M_\odot$ can double their mass and still launch energetic jets.

After the jets activity (the GEE) ends, the system continues to interact via RLOF. This is seen in Fig. \ref{fig:Jet1000p1n2p4Ev} by the changing stellar masses that continues after the GEE ends. At the end it is only the wind that removes part of the left-over envelope of the primary star. The main activity of the GEE is to prevent the CEE, and by that allowing the formation of a SN~IIb progenitor. At the end of the GEE the hydrogen mass is $M_{\rm GEE,H}=4.3 M_\odot$, 
at the end of the final RLOF the hydrogen mass is $M_{\rm RLOF,H}= 0.256 M_\odot$,  
while at core collapse it is $M_{\rm CCSN,H}= 0.058 M_\odot$, fitting a SN~IIb progenitor.  

\begin{figure}
\centering
\includegraphics[trim=0.0cm 4.cm 0.0cm 4.0cm,clip=true,width=0.5\textwidth]{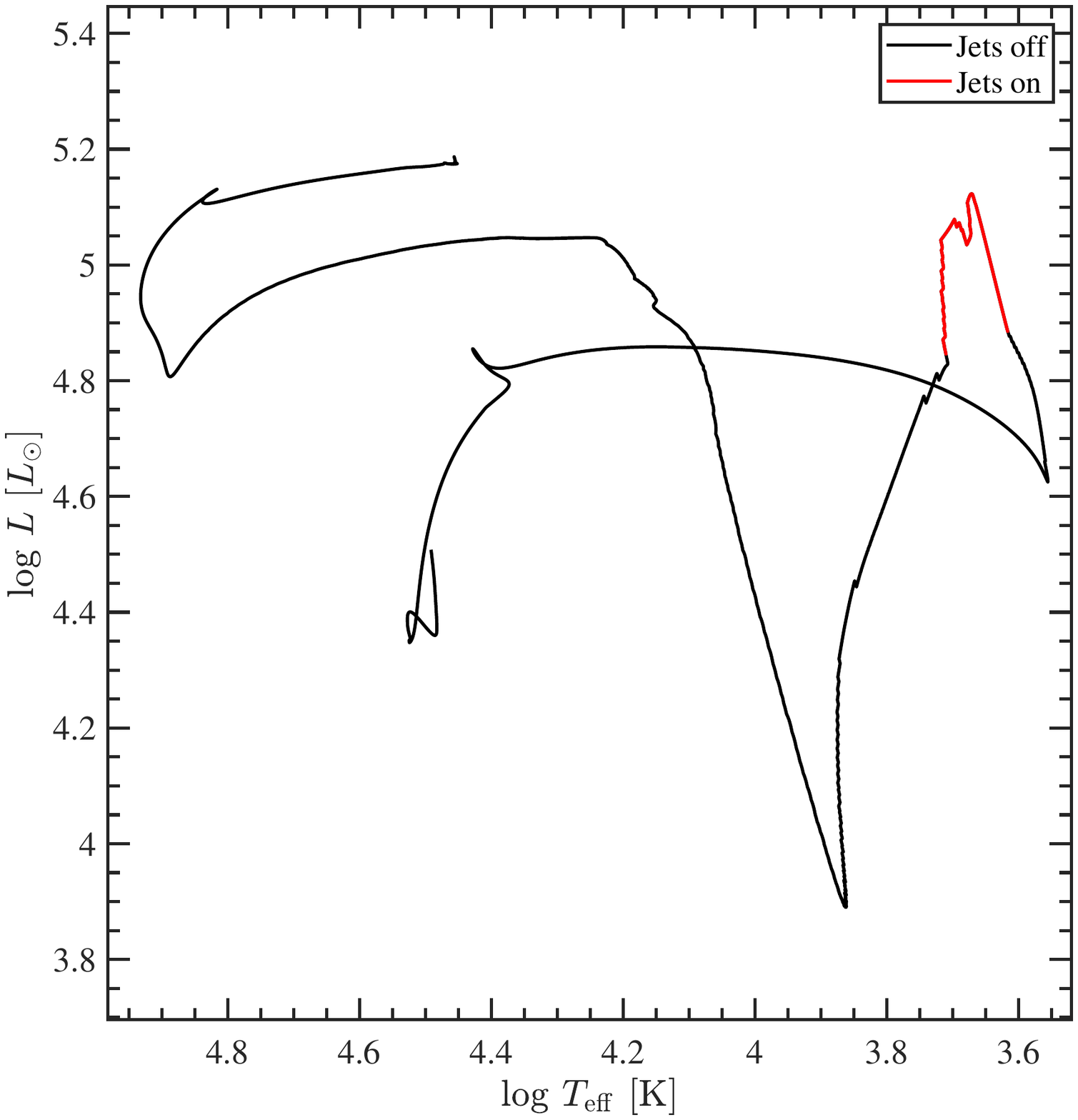}
\vskip -0.0 cm
\caption{Evolution of the primary star in the Jet(1000,1.2,4) case on the HR diagram. The thick-red line marks the activity period of the jets (see also Fig. \ref{fig:Jet1000p1n2p4Ev}). We obtain a blue-compact SN~IIb progenitor with a hydrogen mass of $M_{\rm CCSN,H}= 0.058 M_\odot$ at core collapse.
}
\label{fig:Jet1000p1n2p4HR}
\end{figure}

In Fig. \ref{fig:Jet1000p1n2p4HR} we present the evolution of the primary star of the Jet(1000,1.2,4) case on the HR diagram, from the main sequence to core collapse. The period during which the jets are active, i.e., the GEE, is mark by a thick-red line.  As evident from both Fig. \ref{fig:Jet1000p1n2p4Ev} and Fig. \ref{fig:Jet1000p1n2p4HR} the SN~IIb we obtain here is a blue-compact progenitor (a radius of about $16 R_\odot$) as are most progenitors of SNe~IIb (e.g., \citealt{Yoonetal2017}).   

{{{{ From Figs. \ref{fig:Jet1000p1n2p4Ev} and \ref{fig:Jet1000p1n2p4HR} we learn that mass ejection takes place over a typical time of $\approx 100 \yr$, and that within this time scale the radius of the primary star decreases from $R_1 \simeq 500 R_\odot$ to $R_1 \simeq 50 R_\odot$. The radius continues to decrease over a time scale of $\simeq 5 \times 10^4 \yr$. For about $\approx 10^3-10^5 \yr$ after the enhanced mass loss period there will be a nebula around a luminous blue star. At later times we expect this nebula to disperse in the ISM. Based on 3D hydrodynamical simulations of the GEE  \citep{Shiberetal2017}, 
we expect this nebula to be bipolar, namely, with concentrated equatorial outflow and with faster two opposite wide (tens of degrees) polar outflows.
At early times when the nebula cannot be resolved yet, observations might reveal the bipolar outflow by a highly-polarised emission. }}}}
     
\subsection{Numerical limitations}
\label{subsec:Numerical}

There are several numerical limitations in our usage of the \textsc{mesa~binary} code in relation to mimicking the GEE. The two main problems are the large number of free parameters and the problem in dealing with systems that enter the CEE. 
 
The GEE is a complicated interaction. We here mimic the role of the jets with several free parameters. These are the form of equation (\ref{eq:MLJ1}), the intensity $f_{\rm jet}$, the initial orbital separation of jet activity $R_1 f_{\rm GEE}$, and the fraction of mass that is lost from each star as a result of the jets' activity (here we use half from each star in all runs with jets). 
In addition, there are also the parameters of the RLOF, e.g., $f_{\rm acc,RL}$.
As we have no good handle of these parameters, it is too early to conduct a systematic study of the parameter space. 

The other problem is that the code does not handle well the evolution after the system enters the CEE, and it becomes almost impossible to include our mimicking of the GEE when the secondary star enters the envelope (for that we need full 3D hydrodynamical simulations). We actually expect that in many cases when the secondary star does enter the CEE, the jets, or even the RLOF itself, will be very efficient at removing mass \citep{Shiberetal2017, ShiberSoker2018, LopezCamaraetal2019, Shiberetal2019} and in some cases the system will exit the CEE and will experience the GEE. This is particularly so when the primary rapidly expands and engulfs the secondary star, because in that case the very outer envelope is of very low density. Although in the case we present in section \ref{subsec:prevent} the jets are active only for 4 years, in reality we expect a longer activity while the secondary star enters the very outskirts of the giant envelope (see simulations cited above). 
 
In many cases when we turn on the jets they remove mass from the giant primary star, something that causes the primary star to expand and swallow the secondary star. In most cases we terminate the simulation at this stage for numerical reasons. But as stated, we actually expect the system to experience the GEE along part of the evolution. Some of these binary systems will end with an orbital separation much smaller than the maximum radius that the primary star has achieved along it evolution, similarly to the case we present in Fig. \ref{fig:Jet1000p1n2p4Ev}.  

\begin{figure*}
\centering
\includegraphics[trim=0.0cm 6.0cm 0.0cm 8.0cm,clip=true,width=1.0\textwidth]{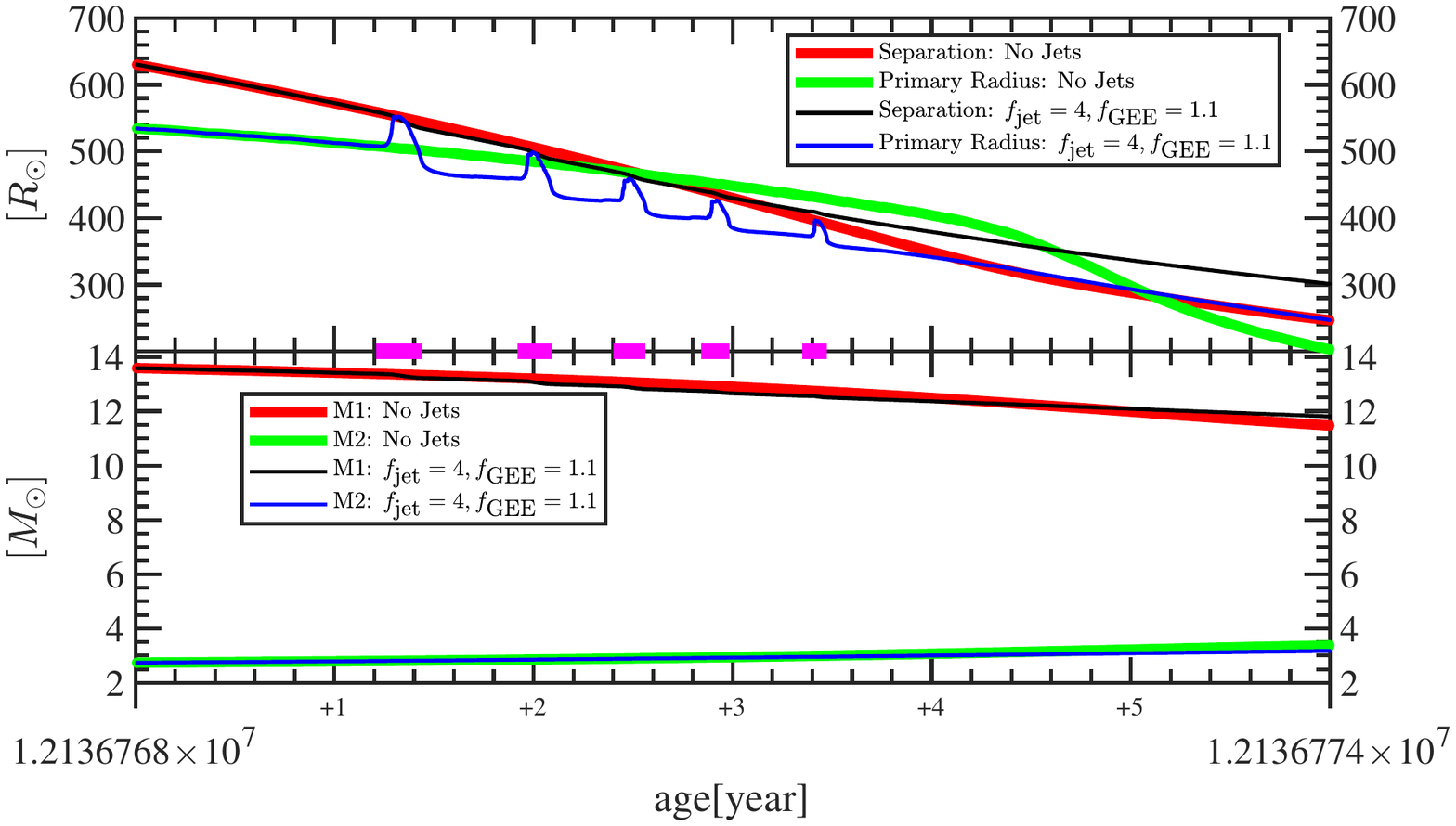}
\vskip -2.0 cm
\caption{Similar to Fig. \ref{fig:Jet1000p1n2p4Ev} but for the Jet(1000,1.1,4) case where the jets are turned on at late time when $a<1.1R_1$ instead of at $a<1.2R_1$. We focus on the relevant time period of the evolution when the strong interaction takes place (middle segment of full evolution similar to that in Fig. \ref{fig:Jet1000p1n2p4Ev}), and continue even after the secondary star enters the envelope, in both cases with and without jets. Note how the primary expands within about a month as a response of mass removal by the jets, and swallows the secondary star. In both cases the system exits the CEE later on, and forms a blue-compact SN~IIb progenitor. }
\label{fig:Jet1000p1n1p4Ev}
\end{figure*}

Because \textsc{mesa~binary} cannot properly treat the CEE, namely the evolution of the secondary star inside the giant envelope, we terminate most simulations when the system does enter a CEE.
Nonetheless, we do present here in Fig. \ref{fig:Jet1000p1n1p4Ev} the evolution of two cases, one with and one without jets, where the system does enter the CEE and then exits the CEE.
We take a case, which we term Jet(1000,1.1,4), where we start the jets late when $a<1.1 R_1$, i.e., $f_{\rm GEE}=1.1$ in equation (\ref{eq:MLJ1}). All other parameters are as in the Jet(1000,1.2,4) case that we present in section \ref{subsec:prevent} and in Fig. \ref{fig:Jet1000p1n2p4Ev}. The case without jets is the same as the NoJet(1000) case that we present there. The results of these runs should be taken with very great caution because \textsc{mesa~binary} cannot properly treat the CEE. However, we do expect that in the GEE the system can get in and out of the CEE, and so the qualitative behaviour might hold, at least for the case with jets \citep{Soker2017}.   

The main results of these runs as evident from Fig. \ref{fig:Jet1000p1n1p4Ev} are that the system can remove enough mass to cause the primary star to shrink and cause the system to detach, i.e., exit the CEE, and eventually form a SN~IIb progenitor. In the case without jets the secondary gets deeper into the envelope (compare thick-green line of primary radius and thick-red line of orbital separation), while in the case with jets the secondary only grazes the envelope at four time periods owing to four expansions of the primary star (compare thin-blue line of primary radius and thin-black line of orbital separation).
This behaviour of several in-and-out phases is expected in some cases of the GEE.  
In both cases the hydrogen mass at core collapse is $M_{\rm CCSN,H}= 0.058 M_\odot$, fitting a SN~IIb progenitor.  

This graph, however, presents another numerical limitation of the scheme we use to mimic the GEE with \textsc{mesa~binary}. The orbital period of the binary system we use here  is $P_{\rm orb}\simeq 0.86 (a/500)^{3/2} \yr$, where $a$ is the orbital separation. 
We can see that the phases of the GEE last each for about $t_{\rm GEE} \approx 1~$month, much shorter than the orbital period. The code treats the primary star as a spherical star, which cannot be the case when $t_{\rm GEE}< P_{\rm orb}$. This is another reason to treat these results with much caution. 

{{{{ We also see in Fig. \ref{fig:Jet1000p1n1p4Ev} that the radius of the primary star suffers `expansion pulses' (the five peaks in the blue line in the upper panel). The sharp behavior of these pulses seems to be a numerical effect, as the numerical code does not yield yet a smooth enough evolution. It seems that the introduction of a dynamical effects in the code might make the behavior more realistic. This should be a separate study, as the introduction of dynamical effects might bring other numerical effects. Eventually, only 3D hydrodynamical simulations, that include non-spherical effects, will be able to capture the entire evolution in a manner closer to reality.  }}}}

We summarise this section as follows. Although the results are very crude for the reasons we listed above, they do suggest that some systems might enter a CEE but then exit the CEE and by RLOF and wind mass removal form SN~IIb progenitors. Although the system in the present case exits the CEE and forms a SN~IIb progenitor even without assuming the operation of jets, the jets seem to ease this processes. Our results strengthen the case for an evolution route where some binary systems enter a CEE, but exit from it to form blue-compact SN~IIb progenitors. 
     
\subsection{Other cases}
\label{subsec:Cases}

We did not conduct a systematic search of the parameter space (see section \ref{subsec:Numerical}). However, we did try 104 cases with and without jets. 
{{{{ We list these cases in Appendix \ref{AppendixA}. }}}}
For the prescription we use with \textsc{mesa~binary} and with the jets-induced mass removal according to equation (\ref{eq:MLJ1}) we have some interesting findings that we describe below, comparing to the Jet(1000,1.2,4), Jet(1000,1.1,4) and NoJet(1000) cases that we present in sections \ref{subsec:prevent} and \ref{subsec:Numerical}. 

If we turn on the jets somewhat earlier than at $a=1.2 R_1$ (i.e., $f_{\rm GEE} > 1.2$) the results do not change much. The jets are active for somewhat a longer time, e.g., about 10-20 years for $f_{\rm GEE}=1.4-1.5$. The star ends with a hydrogen mass at core collapse that is similar to the Jet(1000,1.2,4) case that we present in section \ref{subsec:prevent}. 

In the Jet(1000,1.1,4) case that we present in section \ref{subsec:Numerical} the primary expands as a result of mass removal by jets and for a very short time the system enters a CEE (Fig. \ref{fig:Jet1000p1n1p4Ev}). 
If the jets are weaker, $f_{\rm jet}=2$ instead of $f_{\rm jet}=4$ in run Jet(1000,1.1,4) then the primary expands less and the system avoids a CEE, and ends with a blue-compact SN~IIb progenitor.   

The same effect of rapid primary expansion in response to mass removal by jets occurs when we simulate a system with initial orbital separation of $a_0=1200 R_\odot$ instead of $a_0=1000 R_\odot$. The primary swallows the secondary and we terminate the simulation. Here we also expect that the jets will remove mass even when the secondary is inside the outskirts of the primary envelope, and this case also leads to a SN~IIb progenitor. 
For a case with $a_0=800 R_\odot$ the system does not enter a CEE even without jets because the RLOF starts earlier and removes enough primary mass to prevent CEE. The RLOF followed by the wind, without any jets, reduce the hydrogen mass in the primary star and this leads to the formation of a blue-compact SN~IIb progenitor.

We find also that for a secondary mass of $M_2=3M_\odot$ instead of $M_2=2.5M_\odot$ in the cases that we present in Figs. \ref{fig:Jet1000p1n2p4Ev}-\ref{fig:Jet1000p1n1p4Ev}, even without jets the system avoids a CEE and the RLOF and the later wind remove enough mass to form a blue-compact SN~IIb progenitor. 

\begin{figure*}
\centering
\includegraphics[trim=0.0cm 6.cm 0.0cm 9.0cm,clip=true,width=1.0\textwidth]{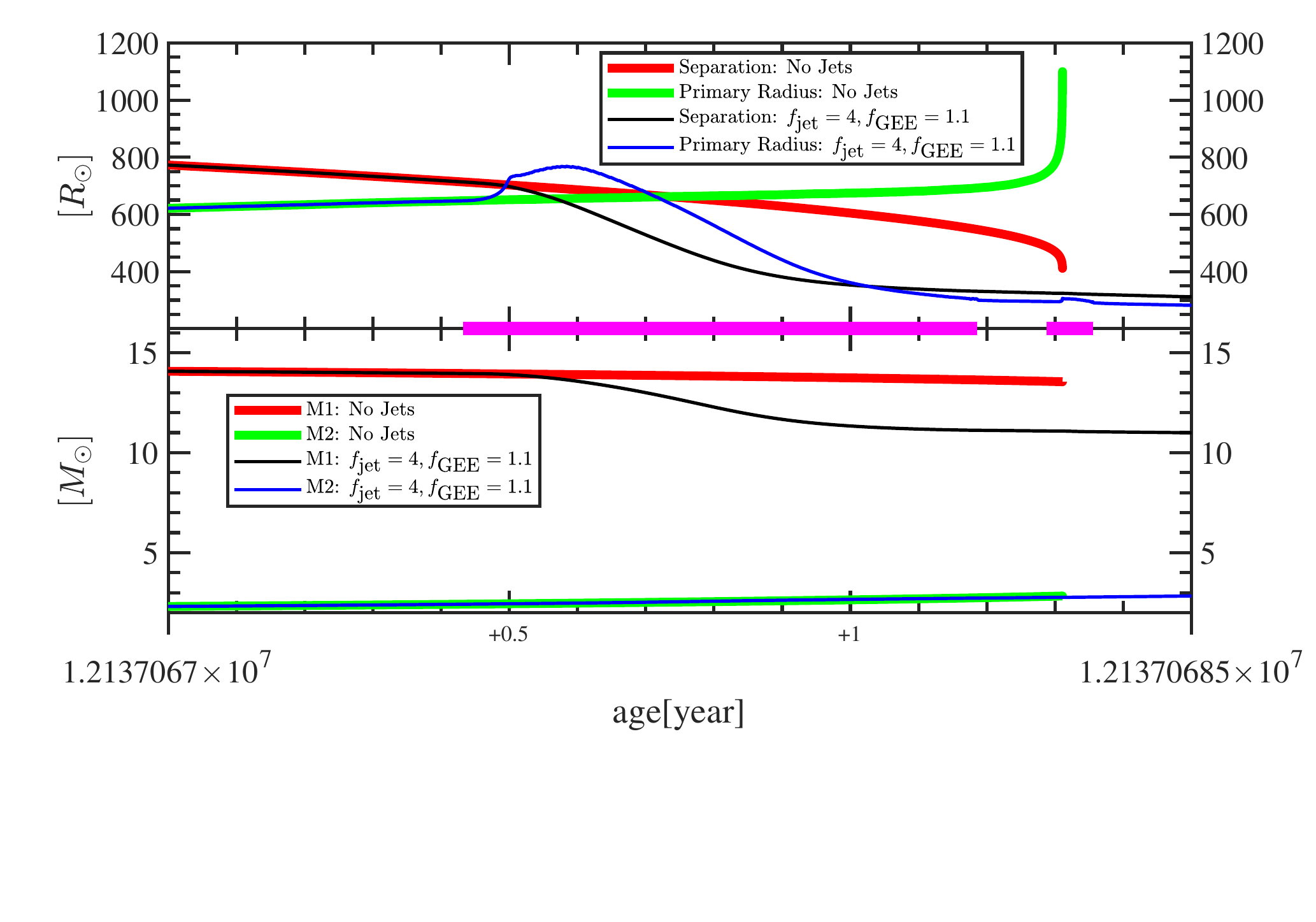}
\vskip -2.0 cm
\caption{Similar to Fig. \ref{fig:Jet1000p1n1p4Ev} but for the Jet(1000,1.1,4,[2,1]) case, that differs from the Jet(1000,1.1,4) case in that the secondary star mass is $M_2=2M_\odot$ instead of $M_2=2.5M_\odot$, and in that the fraction of the RLOF mass that the secondary star accretes is $f_{\rm acc,RL}= 1$ instead of $f_{\rm acc,RL}= 0.3$. In the case with jets the system enters the CEE at an earlier time than the case without jets, but then it exits the CEE to experience a GEE phase, ending with a wind that leaves a hydrogen mass of $M_{\rm CCSN,H}=0.062 M_\odot$, which fits a blue-compact SN~IIb progenitor.  
}
\label{fig:Jet1000p1n5p4EvM2F}
\end{figure*}

In a case with $M_2=2M_\odot$ the evolution to a blue-compact SN~IIb progenitor is different. In Fig. \ref{fig:Jet1000p1n5p4EvM2F} we present the strong interaction time period of the Jet(1000,1.1,4,[2,1]) case, where $a_0=1000 R_\odot$, $f_{\rm GEE}=1.1$ (late jet interaction), $f_{\rm jet}=4$, as in the Jet(1000,1.1,4) case we present in Fig. \ref{fig:Jet1000p1n1p4Ev}, but here the secondary mass is $M_2=2M_\odot$ instead of $2.5 M_\odot$ as in the other figures, and the fraction of the RLOF mass that the secondary star accretes is $f_{\rm acc,RL}= 1$, instead of $f_{\rm acc,RL}= 0.3$ as in all other cases. We also present the case without jets. 

We note the following behaviour in Fig. \ref{fig:Jet1000p1n5p4EvM2F}. 
The jets remove mass from the primary star and this causes its expansion and the early formation of a CEE (thin lines in the figure), i.e., earlier than the case without jets (thick lines). Both in the case with and without jets the initial spiralling-in takes place on a dynamical time of several months, in what is termed the plunge-in phase of the CEE (e.g., \citealt{Ivanovaetal2013}, for a review). Without jets the system continues with the CEE until we terminate the evolution. In the case with jets, on the other hand, the jets remove enough mass and the system exits the CEE and starts a GEE, and then ends with mass removal only by the wind. At core collapse the hydrogen mass is $M_{\rm CCSN,H}=0.062 M_\odot$. 
Namely, in the case with jets the system experiences the GEE and leaves a SN~IIb progenitor. 

For the reasons we discuss in section \ref{subsec:Numerical}, we should take with caution the results we present in Fig. \ref{fig:Jet1000p1n5p4EvM2F}. These reasons include large variations on time-scales shorter or about equal to the dynamical time scale, the inability of the code to treat a non-spherical giant star, and the inaccuracy in the accretion rate when the system enters a CEE. 
The mass removal rate with jets does however make sense under our assumptions.
For $f_{\rm acc,RL}= 1$ equation (\ref{eq:fjet}) gives for the ratio of mass removal rate to mass transfer rate a value of of 10. The maximum ratio we obtain in the Jet(1000,1.1,4,[2,1]) case in our simulation is 16, larger than 10 because $a<R_1$ in equation (\ref{eq:MLJ1}). This makes sense, as when the secondary star enters the envelope the mass accretion rate might be larger than the RLOF.   

The results we present above strengthen the finding of sections \ref{subsec:Numerical} in showing that the GEE can increase the parameter phase for the formation of SN~IIb progenitors. 

{{{{ We end this section with two comments with a broader scope. First we note that our study adds to the importance of the pre-CEE evolution phase in determining the outcome of the CEE itself. Earlier studies pointed out the importance of pre-CEE processes. Examples include the spinning-up process of the giant envelope, a process that allows the giant to lose more of its mass (e.g., \citealt{BearSoker2010}), and the pre-CEE expansion of the giant star that eases unbinding envelope mass (\citealt{Iaconietal2017}). We here included the effects of mass transfer and mass loss in a GEE that might take place during a pre-CEE phase.  }}}}

{{{{ The second point is that the system might exit the GEE with an eccentric orbit. The jet-induced high mass loss rate at periastron passages in the GEE might counteract the circularising effect of the tidal forces. 
\cite{KashiSoker2018} suggest that this effect explains the high eccentricity of post-AGBIBs (see section \ref{subsec:SNIIb}). Here we raise the possibility that the GEE might account for the high eccentricity of some  Wolf-Rayet binary systems, e.g., $\gamma^2$~Velorum that has an eccentricity of $e \simeq 0.33$  \citep{Schmutzetal1997}.  }}}}

\section{SUMMARY}
\label{sec:summary}

We used the stellar evolutionary code \textsc{MESA~binary} to follow the evolution of binary systems that might form SN~IIb progenitors. The new ingredient of the simulations is the introduction of an enhanced mass loss rate due to jets that we assume the secondary star might lunch, according to the assumptions of the GEE (section \ref{subsec:GEE}). In the present study we mimic the jet-driven mass loss according to equation (\ref{eq:MLJ1}).
In section \ref{subsec:Numerical} we discussed some limitations of the jet-driven mass loss prescription we used, and of the numerical code in general. For example, our results when a binary system experiences a CEE should be treated with high caution, as the numerical code does not handle well such a situation. 
In all our simulations the initial mass of the primary star is $M_{\rm 1,i}=15 M_\odot$. We summarise our main findings below.

\begin{enumerate}
\item As other studies have shown (e.g., \citealt{Sravanetal2018}) in some cases RLOF followed by a wind, even if the secondary star does not launch jets, can remove enough mass to form a SN~IIb progenitor. 
\item In some cases where without the effect of jets the system does enter a CEE that might prevent the formation of a SN~IIb progenitor, the jet-driven mass loss might prevent the CEE altogether, as in the case that we present in Fig. \ref{fig:Jet1000p1n2p4Ev}. Because of this jet-driven mass loss, instead of entering a CEE the system experiences a short phase, about four years, of GEE. After that RLOF and then a wind remove most of the envelope mass to leave a blue-compact SN~IIb progenitor (Fig. \ref{fig:Jet1000p1n2p4HR}).
\item In some cases even without jets the system might enter a CEE and exit from it. RLOF and the wind remove enough mass to form a blue-compact SN~IIb progenitor. Short phase(s) of GEE owing to jet-driven mass loss ease this process as we show in Fig. \ref{fig:Jet1000p1n1p4Ev}. \cite{Sravanetal2018} assume that the hydrogen mass at explosion is that when the system enters the CEE. But we found here that in some cases the final hydrogen mass is much lower than that at the onset of the CEE, and can fit that of a SN~IIb progenitor. Namely, this in-and-out of a CEE increases the parameter space of SN~IIb progenitors even before the effects of jets are included.
\item In some cases without jet-driven mass loss the binary system enters and does not exit from a CEE. Unless the system experience a fatal CEE that might lead to a SN~IIb progenitor (\citealt{Lohevetal2019}; section \ref{subsec:SNIIb}), we do not expect that this system forms a SN~IIb progenitor. The introduction of jet-driven mass loss might bring the system to form a SN~IIb progenitor by exiting it from the CEE into a GEE phase, as we show in Fig. \ref{fig:Jet1000p1n5p4EvM2F}. 
\item Overall, our results of the cases we present and 99 other cases (appendix \ref{AppendixA}), show that the process of jet-driven mass loss that leads to episode(s) of GEE phases substantially increases the binary parameter space that leads to the formation of SN~IIb progenitors. This is in accord with the suggestion of \cite{Soker2017} of the GEE scenario for SN~IIb, but some quantitative results are different. For example, we find here the GEE phase to be much shorter than the expectation of \cite{Soker2017}.
In all the cases we have studied the SN~IIb progenitors are blue-compact ones (rather than yellow or red progenitors). 
\end{enumerate}

Our results bring us to try to crudely estimate the fraction of massive CCSN progenitors that experience the GEE.
We proceed as follows. Following our discussion in section \ref{sec:intro} we assume that post-AGB stars with intermediate orbital periods (post-AGBIB stars) are formed by the GEE. We further assume that the post-red giant branch (post-RGB) stars that \cite{Kamathetal2016} study and have similar properties to post-AGBIB stars are also formed by the GEE. \cite{Kamathetal2016} estimate that these post-RGB stars comprise a fraction of 0.0045 of all RGB stars. We also note that the fraction of binary (and higher multiple-stellar systems) of CCSN progenitors is about twice that of solar-like stars (e.g., \citealt{MoeDiStefano2017}). These numbers bring us to conclude that $\approx 1 \%$ of CCSN progenitors experience the GEE under our assumptions. This amounts to $\approx 10 \%$ of SNe~IIb. 
\cite{Sravanetal2018} conclude that at solar metallicty their binary channel can account for $0-2 \%$ of all CCSNe being SNe~IIb. We estimate that the fraction of SN~IIb progenitors that experience the GEE is about equal to that of binary systems that do not experience the GEE. Namely, the GEE might account for $\approx 0-2\%$ of the CCSNe becoming SNe~IIb. 
 
\cite{Soker2019FatalCEE} estimates that the fatal CEE scenario \citep{Lohevetal2019} might account for $1-3 \%$ of all CCSNe. Recent studies suggest that single star channels also contribute to the formation of SNe~IIb (e.g., \citealt{Sravanetal2018}). 

These numbers from the four SN~IIb channels add up to be below the required $\approx 11 \%$ fraction of SNe~IIb out of all CCSNe. We therefore re-scale these numbers to reach the required $\approx 11 \%$ fraction of SNe~IIb out of all CCSNe. Namely, we argue that the different channels must contribute more than what simple (and conservative) estimates (as we listed above) give. This more optimistic estimate is based also on the conclusion (e.g., \citealt{Gilkisetal2019}) that the post-RLOF mass loss rate by wind should be lower than what most studies have assumed, and this increases the number of SN~IIb progenitors by channels that involve RLOF.   
This brings us to suggest, for solar metallicity at least, that each of the following four SN~IIb progenitor channels contributes about the same, that is, each channel contributes $\approx 2-4 \%$ of all CCSNe, to the formation of SNe~IIb: (1) The binary evolution channel with RLOF but without GEE; (2) the GEE; (3) the fatal-CEE; and (4) the single-star channel.

\section*{Acknowledgments}

{{{{ We thank an anonymous referee for helpful comments. }}}} This research was partially supported by the Israel Science Foundation and by a grant from Prof. Amnon Pazy Research Foundation. AG gratefully acknowledges the generous support of the Blavatnik Family Foundation.

\appendix\section{Simulations with \textsc{mesa~binary}}
\label{AppendixA}

In Tables \ref{tabA1} and \ref{tabA2} we summarise the simulations that we have performed with \textsc{mesa~binary}. The first 5 runs are those that we presented in Figs. \ref{fig:Jet1000p1n2p4Ev}-\ref{fig:Jet1000p1n5p4EvM2F} (F1-F4).   
   
   \begin{table*}[]
\begin{tabular}{llllllllll}
Run                       & $M_{\rm 2,i}$ & $a_0$ & $f_{\rm L,RL,1}$ & $f_{\rm L,RL,2}$ & Jets & $f_\mathrm{GEE}$ & $f_{\rm jet}$ & CEE & $M_{\rm H}$      \\
\hline
NoJet(1000) [F1,F3]       & 2.5           & 1000  & 0                & 0.7              & NO      &                  &               & YES & CEE\textbackslash{}T \\
Jet(1000,1.2,4) [F1,F2]   & 2.5           & 1000  & 0                & 0.7              & YES     & 1.2              & 4             & NO  & 0.057749             \\
Jet(1000,1.1,4) [F3]      & 2.5           & 1000  & 0                & 0.7              & YES     & 1.1              & 4             & YES & 0.057749             \\
NoJet(1000[2,1]) [F4]     & 2             & 1000  & 0                & 0                & NO      &                  &               & YES & 5.1598              \\
Jet(1000,1.1,4,[2,1]) [F4] & 2            & 1000  & 0                & 0                & YES     & 1.1              & 4             & YES & 0.061717             \\
1                         & 2.5           & 800   & 0                & 0.7              & NO      &                  &               & YES & CEE\textbackslash{}T \\
2                         & 2.5           & 800   & 0.35             & 0.35             & NO      &                  &               & YES & CEE\textbackslash{}T \\
3                         & 2.5           & 800   & 0.7              & 0                & NO      &                  &               & YES & CEE\textbackslash{}T \\
4                         & 2.5           & 800   & 0                & 0.5              & NO      &                  &               & YES & CEE\textbackslash{}T \\
5                         & 2.5           & 800   & 0.25             & 0.25             & NO      &                  &               & YES & CEE\textbackslash{}T \\
6                         & 2.5           & 800   & 0.5              & 0                & NO      &                  &               & YES & CEE\textbackslash{}T \\
7                         & 2.5           & 800   & 0                & 0.1              & NO      &                  &               & YES & CEE\textbackslash{}T \\
8                         & 2.5           & 800   & 0.05             & 0.05             & NO      &                  &               & YES & CEE\textbackslash{}T \\
9                         & 2.5           & 800   & 0.1              & 0                & NO      &                  &               & YES & CEE\textbackslash{}T \\
10                        & 2.5           & 1000  & 0.7              & 0                & NO      &                  &               & NO  & 0.075488             \\
11                        & 2.5           & 1000  & 0.35             & 0.35             & NO      &                  &               & NO  & 0.069401             \\
12                        & 2.5           & 1000  & 0                & 0.5              & NO      &                  &               & NO  & 0.063522             \\
13                        & 2.5           & 1000  & 0.5              & 0                & NO      &                  &               & NO  & 0.073784             \\
14                        & 2.5           & 1000  & 0.25             & 0.25             & NO      &                  &               & NO  & 0.069522             \\
15                        & 2.5           & 1000  & 0                & 0.9              & NO      &                  &               & YES & CEE\textbackslash{}T \\
16                        & 2.5           & 1000  & 0.9              & 0                & NO      &                  &               & NO  & 0.077908             \\
17                        & 2.5           & 1000  & 0.45             & 0.45             & NO      &                  &               & NO  & 0.06878              \\
18                        & 2.5           & 1000  & 0                & 0.1              & NO      &                  &               & NO  & 0.068973             \\
19                        & 2.5           & 1000  & 0.05             & 0.05             & NO      &                  &               & NO  & 0.069609             \\
20                        & 2.5           & 1000  & 0.1              & 0                & NO      &                  &               & NO  & 0.070201             \\
21                        & 2.5           & 1000  & 0                & 0.7              & YES     & 1.5              & 4             & NO  & 0.057781             \\
22                        & 2.5           & 1000  & 0                & 0.7              & YES     & 1.4              & 4             & NO  & 0.057703             \\
23                        & 2.5           & 1000  & 0                & 0.7              & YES     & 1.3              & 4             & NO  & 0.057729             \\
24                        & 2.5           & 1000  & 0                & 0.7              & YES     & 1.1              & 4             & YES & CEE\textbackslash{}T \\
25                        & 2.5           & 1000  & 0                & 0.7              & YES     & 1.5              & 2             & NO  & 0.0578               \\
26                        & 2.5           & 1000  & 0                & 0.7              & YES     & 1.4              & 2             & NO  & 0.057785             \\
27                        & 2.5           & 1000  & 0                & 0.7              & YES     & 1.3              & 2             & NO  & 0.057736             \\
28                        & 2.5           & 1000  & 0                & 0.7              & YES     & 1.2              & 2             & NO  & 0.057751             \\
29                        & 2.5           & 1000  & 0                & 0.7              & YES     & 1.1              & 2             & NO  & 0.057752             \\
30                        & 2.5           & 1200  & 0                & 0.7              & NO      &                  &               & YES & CEE\textbackslash{}T \\
31                        & 2.5           & 1200  & 0.35             & 0.35             & NO      &                  &               & YES & CEE\textbackslash{}T \\
32                        & 2.5           & 1200  & 0.7              & 0                & NO      &                  &               & YES & CEE\textbackslash{}T \\
33                        & 2.5           & 1200  & 0                & 0.5              & NO      &                  &               & YES & CEE\textbackslash{}T \\
34                        & 2.5           & 1200  & 0.25             & 0.25             & NO      &                  &               & YES & CEE\textbackslash{}T \\
35                        & 2.5           & 1200  & 0.5              & 0                & NO      &                  &               & YES & CEE\textbackslash{}T \\
36                        & 2.5           & 1200  & 0                & 0.1              & NO      &                  &               & YES & CEE\textbackslash{}T \\
37                        & 2.5           & 1200  & 0.05             & 0.05             & NO      &                  &               & YES & CEE\textbackslash{}T \\
38                        & 2.5           & 1200  & 0.1              & 0                & NO      &                  &               & YES & CEE\textbackslash{}T \\
39                        & 2.5           & 1200  & 0                & 0.7              & YES     & 1.5              & 4             & YES & CEE\textbackslash{}T \\
40                        & 2.5           & 1200  & 0                & 0.7              & YES     & 1.4              & 4             & YES & CEE\textbackslash{}T \\
41                        & 2.5           & 1200  & 0                & 0.7              & YES     & 1.3              & 4             & YES & CEE\textbackslash{}T \\
42                        & 2.5           & 1200  & 0                & 0.7              & YES     & 1.2              & 4             & YES & CEE\textbackslash{}T \\
43                        & 2.5           & 1200  & 0                & 0.7              & YES     & 1.1              & 4             & YES & CEE\textbackslash{}T \\
44                        & 2.5           & 1200  & 0                & 0.7              & YES     & 1.5              & 2             & YES & CEE\textbackslash{}T \\
45                        & 2.5           & 1200  & 0                & 0.7              & YES     & 1.4              & 2             & YES & CEE\textbackslash{}T \\
46                        & 2.5           & 1200  & 0                & 0.7              & YES     & 1.3              & 2             & YES & CEE\textbackslash{}T \\
47                        & 2.5           & 1200  & 0                & 0.7              & YES     & 1.2              & 2             & YES & CEE\textbackslash{}T \\
48                        & 2.5           & 1200  & 0                & 0.7              & YES     & 1.1              & 2             & YES & CEE\textbackslash{}T \\
\hline
\end{tabular}
\caption{The different cases that we simulated, all for a primary initial mass of $M_{\rm 1,i}=15 M_\odot$. The first five runs are those that we present in the figures (F1,F2,F3 and F4). Runs 1-48 are cases with $M_2=2.5 M_\odot$. 
The different columns are as follows. 
$M_{\rm 2,i}$ is the initial secondary mass in $M_\odot$; $a_0$ the initial orbital separation in $R_\odot$; 
$f_{\rm L,RL,1}$ is the fraction of the mass that the primary star loses as a result of RLOF and that leaves the system directly from the primary; $f_{\rm L,RL,2}$ is the fraction of the mass that the primary star loses as a result of RLOF and that leaves the system through the secondary star $M_2$; `Jets' indicates whether we allow for the enhanced mass loss rate due to jets (equation \ref{eq:MLJ1}) to operate when condition (\ref{eq:fGEE}) holds (YES), or whether we do not allow for the effect of jets (NO); $f_\mathrm{GEE}$ is the jet-activity separation factor as defined in equation (\ref{eq:fGEE});  $f_{\rm jet}$ is the jet-driven mass loss factor as defined in equation (\ref{eq:MLJ1}); CEE indicates whether the system enters (YES) or does not enter (NO) a CEE phase. $M_{\rm H}$ is the hydrogen mass at the end of the simulation. In all cases with $M_{\rm H} < 0.1 M_\odot$ this is the mass at core collapse (explosion)  $M_{\rm CCSN, H}$.  
In some cases we continued the simulations into the CEE. In most cases we terminated the evolution when the system entered the CEE, and these are marked with `CEE\textbackslash{}T'.  
}
\label{tabA1}
\end{table*}

   \begin{table*}[]
\begin{tabular}{llllllllll}
Run                       & $M_{\rm 2,i}$ & $a_0$ & $f_{\rm L,RL,1}$ & $f_{\rm L,RL,2}$ & Jets & $f_\mathrm{GEE}$ & $f_{\rm jet}$ & CEE & $M_{\rm H}$      \\
\hline
49                        & 1.5           & 1000  & 0                & 0.7              & NO      &                  &               & YES & CEE\textbackslash{}T \\
50                        & 1.5           & 1000  & 0                & 0.7              & YES     & 1.5              & 4             & YES & CEE\textbackslash{}T \\
51                        & 1.5           & 1000  & 0                & 0.7              & YES     & 1.4              & 4             & YES & CEE\textbackslash{}T \\
52                        & 1.5           & 1000  & 0                & 0.7              & YES     & 1.3              & 4             & YES & CEE\textbackslash{}T \\
53                        & 1.5           & 1000  & 0                & 0.7              & YES     & 1.2              & 4             & YES & CEE\textbackslash{}T \\
54                        & 1.5           & 1000  & 0                & 0.7              & YES     & 1.1              & 4             & YES & CEE\textbackslash{}T \\
55                        & 2             & 1000  & 0                & 0.7              & NO      &                  &               & YES & CEE\textbackslash{}T \\
56                        & 2             & 1000  & 0                & 0.7              & NO      &                  &               & YES & 5.1422               \\
57                        & 2             & 1000  & 0                & 0.7              & YES     & 1.5              & 4             & YES & 4.1969               \\
58                        & 2             & 1000  & 0                & 0.7              & YES     & 1.1              & 4             & YES & 0.046895             \\
59                        & 2             & 1000  & 0                & 0.7              & YES     & 1.5              & 4             & YES & CEE\textbackslash{}T \\
60                        & 2             & 1000  & 0                & 0.7              & YES     & 1.4              & 4             & YES & CEE\textbackslash{}T \\
61                        & 2             & 1000  & 0                & 0.7              & YES     & 1.3              & 4             & YES & CEE\textbackslash{}T \\
62                        & 2             & 1000  & 0                & 0.7              & YES     & 1.2              & 4             & YES & CEE\textbackslash{}T \\
63                        & 2             & 1000  & 0                & 0.7              & YES     & 1.1              & 4             & YES & CEE\textbackslash{}T \\
64                        & 2             & 1000  & 0                & 0                & YES     & 1.5              & 4             & YES & 0.061055             \\
65                        & 3             & 1000  & 0                & 0.7              & NO      &                  &               & NO  & 0.064111             \\
66                        & 3             & 800   & 0                & 0.7              & NO      &                  &               & NO  & 0.061204             \\
67                        & 3             & 800   & 0.35             & 0.35             & NO      &                  &               & NO  & 0.07023              \\
68                        & 3             & 800   & 0.7              & 0                & NO      &                  &               & NO  & 0.074899             \\
69                        & 3             & 800   & 0                & 0.5              & NO      &                  &               & NO  & 0.065527             \\
70                        & 3             & 800   & 0.25             & 0.25             & NO      &                  &               & NO  & 0.070061             \\
71                        & 3             & 800   & 0.5              & 0                & NO      &                  &               & NO  & 0.073421             \\
72                        & 3             & 800   & 0                & 0.1              & NO      &                  &               & NO  & 0.069385             \\
73                        & 3             & 800   & 0.05             & 0.05             & NO      &                  &               & NO  & 0.069859             \\
74                        & 3             & 800   & 0.1              & 0                & NO      &                  &               & NO  & 0.070314             \\
75                        & 3             & 600   & 0                & 0.7              & NO      &                  &               & NO  & 0.057075             \\
76                        & 3             & 600   & 0.35             & 0.35             & NO      &                  &               & NO  & 0.067469             \\
77                        & 3             & 600   & 0.7              & 0                & NO      &                  &               & NO  & 0.074379             \\
78                        & 3             & 600   & 0                & 0.5              & NO      &                  &               & NO  & 0.061715             \\
79                        & 3             & 600   & 0.25             & 0.25             & NO      &                  &               & NO  & 0.067259             \\
80                        & 3             & 600   & 0.5              & 0                & NO      &                  &               & NO  & 0.071372             \\
81                        & 3             & 600   & 0                & 0.1              & NO      &                  &               & NO  & 0.06641              \\
82                        & 3             & 600   & 0.05             & 0.05             & NO      &                  &               & NO  & 0.067017             \\
83                        & 3             & 600   & 0.1              & 0                & NO      &                  &               & NO  & 0.067591             \\
84                        & 2.1           & 1000  & 0                & 0.7              & NO      &                  &               & YES & CEE\textbackslash{}T \\
85                        & 2.2           & 1000  & 0                & 0.7              & NO      &                  &               & YES & CEE\textbackslash{}T \\
86                        & 2.3           & 1000  & 0                & 0.7              & NO      &                  &               & YES & CEE\textbackslash{}T \\
87                        & 2.4           & 1000  & 0                & 0.7              & NO      &                  &               & YES & CEE\textbackslash{}T \\
88                        & 2.6           & 1000  & 0                & 0.7              & NO      &                  &               & NO  & 0.059119             \\
89                        & 2.7           & 1000  & 0                & 0.7              & NO      &                  &               & NO  & 0.060333             \\
90                        & 2.8           & 1000  & 0                & 0.7              & NO      &                  &               & NO  & 0.061753             \\
91                        & 2.9           & 1000  & 0                & 0.7              & NO      &                  &               & NO  & 0.06301              \\
92                        & 2.1           & 1000  & 0                & 0.7              & YES     & 1.5              & 4             & YES & CEE\textbackslash{}T \\
93                        & 2.2           & 1000  & 0                & 0.7              & YES     & 1.5              & 4             & YES & CEE\textbackslash{}T \\
94                        & 2.3           & 1000  & 0                & 0.7              & YES     & 1.5              & 4             & NO  & 0.053711             \\
95                        & 2.4           & 1000  & 0                & 0.7              & YES     & 1.5              & 4             & NO  & 0.056133             \\
96                        & 2.1           & 1000  & 0                & 0.7              & YES     & 1.5              & 2             & YES & CEE\textbackslash{}T \\
97                        & 2.2           & 1000  & 0                & 0.7              & YES     & 1.5              & 2             & YES & CEE\textbackslash{}T \\
98                        & 2.3           & 1000  & 0                & 0.7              & YES     & 1.5              & 2             & NO  & 0.053842             \\
99                        & 2.4           & 1000  & 0                & 0.7              & YES     & 1.5              & 2             & NO  & 0.056224 \\
\hline
\end{tabular}
\caption{Like Table \ref{tabA1} but for $M_2 \ne 2.5 M_\odot$.  } 
\label{tabA2}
\end{table*}

\label{lastpage}
\end{document}